\documentclass{e1}
\usepackage{amssymb}
\usepackage{rotating}
\usepackage{longtable}
\usepackage{graphicx}
\usepackage{natbib}
\begin{document}
\begin{frontmatter}

\title{CODON USAGE BIAS MEASURED THROUGH ENTROPY APPROACH}
\author[ibf,defakto]{Michael G.Sadovsky\corauthref{cor1}\thanksref{label2}}
\thanks[label2]{To whom the correspondence should be addressed.}
\corauth[cor1]{660036 Russia, Krasnoyarsk, Akademgorodok; Institute of computational modelling of RAS; tel.
+7(3912)907469, fax: +7(3912)907454}
\address[ibf]{Institute of computational modelling of RAS}
\ead{msad@icm.krasn.ru}
\author[defakto]{Julia A.Putintzeva}
\address[defakto]{Siberian Federal university, Institute of natural sciences \& humanities}
\ead{kinomanka85@mail.ru}

\begin{abstract}
Codon usage bias measure is defined through the mutual entropy calculation of real codon frequency
distribution against the quasi-equilibrium one. This latter is defined in three manners: (1) the frequency of
synonymous codons is supposed to be equal (i.e., the arithmetic mean of their frequencies); (2) it coincides
to the frequency distribution of triplets; and, finally, (3) the quasi-equilibrium frequency distribution is
defined as the expected frequency of codons derived from the dinucleotide frequency distribution. The measure
of bias in codon usage is calculated for $125$ bacterial genomes.
\end{abstract}


\end{frontmatter}

\newpage
\section{Introduction}\label{intro}

It is a common fact, that the genetic code is degenerated. All amino acids (besides two ones) are encoded by
two or more codons; such codons are called synonymous and usually differ in a nucleotide occupying the third
position at codon. The synonymous codons occur with different frequencies, and this difference is observed
both between various genomes \citep{1,2,3,4}, and different genes of the same genome \citep{3,4,5,6}. A
synonymous codon usage bias could be explained in various ways, including mutational bias (shaping genomic
$\mathsf{G}$+$\mathsf{C}$ composition) and translational selection by tRNA abundance (acting mainly on highly
expressed genes). Still, the reported results are somewhat contradictory \citep{6}. A contradiction may
result from the differences in statistical methods used to estimate the codon usage bias. Here one should
clearly understand what factors affect the method and numerical result. Boltzmann entropy theory
\citep{bolz,e2} has been applied to estimate the degree of deviation from equal codon usage \citep{x,3}.

The key point here is that the deviation measure of codon usage bias should be independent of biological
issue. It is highly desirable to avoid an implementation of any biological assumptions (such as mutational
bias or translational selection); it must be defined in purely mathematical way. The idea of entropy seems to
suit best of all here. The additional constraints on codon usage resulted from the amino acid frequency
distribution affects the entropy values, thus conspiring the effects directly linked to biases in synonymous
codon usage.

Here we propose three new indices of codon usage bias, which take into account all of the three important
aspects of amino acid usage, i.e. (1) the number of distinct amino acids, (2) their relative frequencies, and
(3) their degree of codon degeneracy. All the indices are based on mutual entropy $\overline{S}$ calculation.
They differ in the codon frequency distribution supposed to be ``quasi-equilibrium". Indeed, the difference
between the indices consists in the difference of the definition of that latter.

Consider a genetic entity, say, a genome, of the length $N$; that latter is the number of nucleotides
composing the entity. A word $\omega$ (of the length $q$) is a string of the length $q$, $1 \leq q \leq N$
observed within the entity. A set of all the words occurred within an entity makes the support $\mathsf{V}$
of the entity (or $q$--support, if indication of the length $q$ is necessary). Accompanying each element
$\omega$, $\omega \in \mathsf{V}$ with the number $n_{\omega}$ of its copies, one gets the (finite)
dictionary of the entity. Changing $n_{\omega}$ for the frequency \[f_{\omega}= \frac{n_{\omega}}{N}\,,\] one
gets the frequency dictionary $W_q$ of the entity (of the thickness $q$).

Everywhere below, for the purposes of this paper, we shall distinguish codon frequency distribution from the
triplet frequency distribution. A triplet frequency distribution is the frequency dictionary $W_3$ of the
thickness $q=3$, where triplets are identified with neither respect to the specific position of a triplet
within the sequence. On the contrary, codon distribution is the frequency distribution of the triplets
occupying specific places within an entity: a codon is the triplet embedded into a sequence at the coding
position, only. Thus, the abundance of copes of the words of the length $q=3$ involved into the codon
distribution implementation is three times less, in comparison to the frequency dictionary $W_3$ of triplets.
Further, we shall denote the codon frequency dictionary as $\mathfrak{W}$; no lower index will be used, since
the thickness of the dictionary is fixed (and equal to $q=3$).

\section{Materials and methods}\label{sec:1}
\subsection{Sequences and Codon Tabulations}\label{sec:2}
The tables of codon usage frequency were taken at Kazusa Institute site\footnote{www.kazusa.ac.jp/codons}.
The corresponding genome sequences have been retrieved from EMBL--bank\footnote{www.ebi.ac.uk/genomes}. The
codon usage tables containing not less that $10000$ codons have been used. Here we studied bacterial genomes
(see Table~\ref{T1}).

\subsection{Codon bias usage indices}\label{sec:2-2}
Let $F$ denote the codon frequency distribution, $F = \{f_{\nu_1\nu_2\nu_3}\}$; here $f_{\nu_1\nu_2\nu_3}$ is
the frequency of a codon $\nu_1\nu_2\nu_3$. Further, let $\widetilde{F}$ denote a quasi-equilibrium frequency
distribution of codons. Hence, the measure $I$ of the codon usage bias is defined as the mutual entropy of
the real frequency distribution $F$ calculated against the quasi-equilibrium $\widetilde{F}$ one:
\begin{equation}\label{eq:1}
I = \sum_{\omega = 1}^{64} f_{\omega} \cdot \ln \left( \frac{f_{\omega}}{\tilde{f}_{\omega}} \right)\;.
\end{equation}
Here index $\omega$ enlists the codons, and $\tilde{f}_{\omega} \in \widetilde{F}$ is quasi-equilibrium
frequency. The measure (\ref{eq:1}) itself is rather simple and clear; a definition of quasi-equilibrium
distribution of codons is the matter of discussion here. We propose three ways to define the distribution
$\widetilde{F}$; they provide three different indices of codon usage bias. The relation between the values of
these indices observed for the same genome is the key issue, for our study.

\subsubsection{Locally equilibrium codon distribution}
It is well known fact, that various amino acids manifest different occurrence frequency, within a genome, or
a gene. Synonymous codons, in turn, exhibit the different occurrence within the similar genetic entities.
Thus, an equality of frequencies of all the synonymous codons encoding the same amino acid
\begin{equation}\label{eq:2}
\tilde{f}_j = \frac{1}{L} \sum_{j \in J_i} f_j\,, \qquad \sum_{j \in J_i} f_j = \sum_{j \in J_i} \tilde{f}_j
= \varphi_i \;,
\end{equation}
is the first way to determine a quasi-equilibrium codon frequency distribution. Here the index $j$ enlists
the synonymous codons encoding the same amino acid, and $J_i$ is the set of such codons for $i{\textrm{-th}}$
amino acid, and $\varphi_i$ is the frequency of that latter. Surely, the list of amino acids must be extended
with {\sl stop} signal (encoded by three codons). Obviously, $\tilde{f}_j = \tilde{f}_k$ for any couple $j,k
\in J_i$.

\subsubsection{Codon distribution vs. triplet distribution}
A triplet distribution gives the second way to define the quasi-equilibrium codon frequency distribution.
Since the codon frequency is determined with respect to the specific locations of the strings of the length
$q=3$, then two third of the abundance of copies of these strings fall beyond the calculation of the codon
frequency distribution. Thus, one can compare the codon frequency distribution with the similar distribution
implemented over the entire sequence, with no gaps in strings location. So, the frequency dictionary of the
thickness $q=3$
\begin{equation}\label{eq:3}
\tilde{f}_l = \hat{f}_l\,, \qquad 1 \leq l \leq 64
\end{equation}
is the quasi-equilibrium codon distribution here.

\subsubsection{The most expected codon frequency distribution}
Finally, the third way to define the quasi-equilibrium codon frequency distribution is to derive it from the
frequency distribution of dinucleotides composing the codon. Having the codons frequency distribution $F$,
one always can derive the frequency composition $F_2$ of the dinucleotides composing the codons. To do that,
one must sum up the frequencies of the codons differing in the third (or the first one) nucleotide. Such
transformation is unambiguous\footnote{Here one must close up a sequence into a ring.}. The situation is
getting worse, as one tends to get a codon distribution due to the inverse transformation. An upward
transformation yields a family of dictionaries $\{F\}$, instead of the single one $F$. To eliminate the
ambiguity, one should implement some basic principle in order to avoid an implementation of extra, additional
information into the codon frequency distribution development. The principle of maximum of entropy of the
extended (i.e., codon) frequency distribution makes sense here \citep{n1,n2,n3,n4}. It means that a
researcher must figure out the extended (or reconstructed) codon distribution $\widetilde{F}$ with maximal
entropy, among the entities composing the family $\{F\}$. This approach allows to calculate the frequencies
of codons explicitly:
\begin{equation}\label{eq:4}
\widetilde{f}_{ijk} = \frac{f_{ij}\times f_{jk}}{f_{j}}\;,
\end{equation}
where $\widetilde{f}_{ijk}$ is the expected frequency of codon $ijk$, $f_{ij}$ is the frequency of a
dinucleotide $ij$, and $f_j$ is the frequency of nucleotide $j$; here $i,j,k \in \{\mathsf{A}, \mathsf{C},
\mathsf{G}, \mathsf{T}\}$.

Thus, the calculation of the measure (\ref{eq:1}) maps each genome into tree-dimension space. Table~\ref{T1}
shows the data calculated for 115 bacterial genomes.

\section{Results}\label{res}
We have examined 115 bacterial genomes. The calculations of three indices (\ref{eq:1}~-- \ref{eq:4}) and the
absolute entropy of codon distribution is shown in Table~\ref{T1}.
\begin{longtable}{|p{8.4cm}|c|c|c|c|c|}
\caption{\label{T1} Indices of codon usage bias; is the index calculated according to (\ref{eq:2}),
$S^{\ast}$ stands for the index defined due to (\ref{eq:3}),
and $T$ is the index defined due to (\ref{eq:4}). $S$ is the absolute entropy of codon distribution. $C$ is the class attribution (see Section~\ref{classif}).}\\
\hline \multicolumn{1}{|c|}{Genomes}& \multicolumn{1}{c|}{$I$} & $S^{\ast}$ & $T$ & $S$ & $C$\\
\hline
\endfirsthead
\multicolumn{6}{r}%
{{\tablename\ \thetable{} -- continued}} \\
\hline \multicolumn{1}{|c|}{Genomes} & \multicolumn{1}{c|}{$I$} & $S^{\ast}$ & $T$ & $S$ & $C$\\
\hline
\endhead
\hline \multicolumn{6}{|r|}{{continued on the next page}} \\ \hline
\endfoot
\endlastfoot
Acinetobacter sp.ADP1&0.1308&0.1526&0.1332&3.9111&1\\
Aeropyrum pernix K1&0.1381&0.1334&0.1611&3.9302&2\\
Agrobacterium tumefaciens str. C58&0.1995&0.1730&0.2681&3.8504&2\\
Aquifex aeolicus VF5&0.1144&0.1887&0.2273&3.8507&2\\
Archaeoglobus fulgidus DSM 4304&0.1051&0.2008&0.2264&3.9011&2\\
Bacillus anthracis str. Ames&0.1808&0.1880&0.1301&3.8232&1\\
Bacillus anthracis str. Sterne&0.1800&0.1873&0.1300&3.8236&1\\
Bacillus anthracis str.'Ames Ancestor'&0.1788&0.1850&0.1278&3.8246&1\\
Bacillus cereus ATCC 10987&0.1750&0.1791&0.1254&3.8291&1\\
Bacillus cereus ATCC 14579&0.1807&0.1853&0.1290&3.8220&1\\
Bacillus halodurans C-125&0.0538&0.1296&0.0967&3.9733&1\\
Bacillus subtilis subsp.subtilis str. 168&0.0581&0.1231&0.1117&3.9605&2\\
Bacteroides fragilis YCH46&0.0499&0.1201&0.1305&3.9824&2\\
Bacteroides thetaiotaomicron VPI-5482&0.0557&0.1258&0.1364&3.9713&2\\
Bartonella henselae str. Houston-1&0.1555&0.1650&0.1077&3.8913&1\\
Bartonella quintana str. Toulouse&0.1525&0.1616&0.1039&3.8954&1\\
Bdellovibrio bacteriovorus HD100&0.1197&0.1593&0.2404&3.9232&2\\
Bifidobacterium longum NCC2705&0.2459&0.2315&0.3666&3.8011&2\\
Bordetella bronchiseptica RB50&0.4884&0.3165&0.5598&3.5485&2\\
Borrelia burqdorferi B31&0.2330&0.1555&0.0988&3.6709&1\\
Borrelia garinii Pbi&0.2421&0.1616&0.1008&3.6630&1\\
Bradyrhizobium japonicum USDA 110&0.3163&0.2236&0.3789&3.7368&2\\
Campylobacter jejuni RM1221&0.2839&0.1994&0.1357&3.6617&1\\
Campylobacter jejuni subsp. Jejuni NCTC 11168&0.2846&0.2010&0.1379&3.6660&1\\
Caulobacter crescentus CB15&0.4250&0.2890&0.5045&3.6062&2\\
Chlamydophila caviae GPIC&0.1079&0.1199&0.0990&3.9445&1\\
Chlamydophila pneumoniae CWL029&0.0803&0.1054&0.0778&3.9748&1\\
Chlamydophila pneumoniae J138&0.0801&0.1050&0.0772&3.9755&1\\
Chlamydophila pneumoniae TW-183&0.0802&0.1037&0.0764&3.9760&1\\
Chlorobium tepidum TLS&0.1767&0.1809&0.2935&3.8777&2\\
Chromobacterium violaceum ATCC 12472&0.4245&0.3004&0.5354&3.6218&2\\
Clamydophyla pneumoniae AR39&0.0804&0.1055&0.0773&3.9748&2\\
Clostridium acetobutylicum ATCC 824&0.2431&0.1951&0.1305&3.7142&1\\
Clostridium perfringens str. 13&0.3602&0.2752&0.1943&3.5816&1\\
Clostridium tetani E88&0.3240&0.2381&0.1767&3.6088&1\\
Corynebacterium efficiens YS-314&0.2983&0.2379&0.3980&3.7494&2\\
Corynebacterium glutamicum ATCC 13032&0.0964&0.1510&0.1674&3.9498&2\\
Coxiella burnetii RSA 493&0.0843&0.1050&0.0892&3.9648&2\\
Desulfovibrio vulgaris subsp.vulgaris str. Hildenborough&0.2459&0.1980&0.3183&3.8090&2\\
Enterococcus faecalis V583&0.1592&0.1838&0.1295&3.8453&1\\
Escherichia coli CFT073&0.1052&0.1305&0.1734&3.9576&2\\
Escherichia coli K12 MG1655&0.1206&0.1463&0.1933&3.9372&2\\
Helicobacter hepaticus ATCC 51449&0.1760&0.1513&0.1065&3.8315&1\\
Helicobacter pylori 26695&0.1420&0.1646&0.1843&3.8454&2\\
Helicobacter pylori J99&0.1404&0.1660&0.1895&3.8479&2\\
Lactobacillus johnsonii NCC 533&0.2113&0.1937&0.1481&3.7856&1\\
Lactobacillus plantarum WCFS1&0.0813&0.1453&0.1544&3.9537&2\\
Lactococcus lactis subsp. Lactis Il1403&0.1923&0.1857&0.1173&3.8068&1\\
Legionella pneumophila subsp. Pneumophila str. Philadelphia 1&0.1018&0.1098&0.0880&3.9339&1\\
Leifsonia xyli subsp. Xyli str. CTCB07&0.3851&0.2411&0.4032&3.6490&2\\
Listeria monocytoqenes str. 4b F2365&0.1389&0.1766&0.1012&3.8600&1\\
Mannheimia succiniciproducens MBEL55E&0.1390&0.1624&0.1571&3.8943&1\\
Mesorhizobium loti MAFF303099&0.2734&0.2019&0.3402&3.7751&2\\
Methanocaldococcus jannaschii DSM 2661&0.2483&0.2108&0.1324&3.6751&2\\
Methanopyrus kandleri AV19&0.2483&0.2108&0.1324&3.6751&1\\
Methanosarcina acetivorans C2A&0.0530&0.1223&0.0876&3.9718&1\\
Methanosarcina mazei Go1&0.0739&0.1314&0.0889&3.9468&1\\
Methylococcus capsulatus str. Bath&0.2847&0.2096&0.3738&3.7709&2\\
Mycobacterium avium subsp. Paratuberculosis str. K10&0.4579&0.2779&0.4819&3.6038&2\\
Mycobacterium bovis AF2122/97&0.2449&0.1688&0.2862&3.7931&2\\
Mycobacterium leprae TN&0.1075&0.1216&0.1717&3.9513&2\\
Mycobacterium tuberculoisis CDC1551&0.2387&0.1618&0.2749&3.8029&2\\
Mycobacterium tuberculosis H37Rv&0.2457&0.1696&0.2878&3.7929&2\\
Mycoplasma mycoides subsp. mycoides SC&0.4748&0.2571&0.2247&3.4356&1\\
Mycoplasma penetrans HF-2&0.4010&0.2320&0.2047&3.5294&1\\
Neisseria gonorrhoeae FA 1090&0.1610&0.1740&0.2343&3.8852&2\\
Neisseria meningitidis MC58&0.1481&0.1708&0.2244&3.8969&2\\
Neisseria meninqitidis Z2491 serogroup A str. Z2491&0.1541&0.1786&0.2342&3.8898&2\\
Nitrosomonas europeae ATCC 19718&0.0824&0.1104&0.1587&3.9806&2\\
Nocardia farcinica IFM 10152&0.4842&0.2917&0.4968&3.5343&2\\
Nostoc sp.PCC7120&0.0877&0.1308&0.1124&3.9638&1\\
Parachlamydia sp. UWE25&0.1689&0.1397&0.1027&3.8561&1\\
Photorhabdus luminescens subsp. Laumondii TTO1&0.0704&0.1183&0.1068&3.9838&1\\
Porphyromonas gingivalis W83&0.0476&0.1167&0.1559&4.0034&2\\
Prochlorococcus marinus str. MIT 9313&0.0472&0.0956&0.0773&4.0203&1\\
Prochlorococcus marinus subsp. Marinus str. CCMP1375&0.1729&0.1423&0.1177&3.8697&1\\
Prochlorococcus marinus subsp. Pastoris str. CCMP1986&0.2556&0.1671&0.1412&3.7354&1\\
Propionibacterium acnes KPA171202&0.1277&0.1338&0.1700&3.9293&2\\
Pseudomonas aeruginosa PAO1&0.4648&0.3204&0.5733&3.5827&2\\
Pseudomonas putida KT2440&0.2847&0.2255&0.4061&3.7696&2\\
Pseudomonas syringae pv. Tomato str. DC3000&0.1960&0.1736&0.3013&3.8633&2\\
Pyrococcus abyssi GE5&0.0983&0.1962&0.1996&3.8887&2\\
Pyrococcus furiosus DSM 3638&0.1000&0.1641&0.1079&3.8847&1\\
Pyrococcus horikoshii OT3&0.0899&0.1508&0.1260&3.9105&1\\
Salmonella enterica subsp. Enterica serovar Typhi Ty2&0.1272&0.1465&0.2068&3.9327&2\\
Salmonella typhimurium LT2&0.1293&0.1490&0.2100&3.9300&2\\
Shewanella oneidensis MR-1&0.0700&0.1320&0.1329&3.9795&2\\
Shigella flexneri 2a str. 2457T&0.1196&0.1429&0.1913&3.9416&2\\
Shigella flexneri 2a str. 301&0.1097&0.1343&0.1791&3.9529&2\\
Sinorhizobium meliloti 1021&0.1960&0.2199&0.3013&3.8633&2\\
Staphylococcus aureus subsp. Aureus MRSA252&0.2338&0.2086&0.1531&3.7572&1\\
Staphylococcus aureus subsp. Aureus MSSA476&0.2356&0.2071&0.1554&3.7557&1\\
Staphylococcus aureus subsp. Aureus Mu50&0.2318&0.2056&0.1522&3.7591&1\\
Staphylococcus aureus subsp. Aureus MW2&0.2368&0.2106&0.1562&3.7535&1\\
Staphylococcus aureus subsp. Aureus N315&0.2348&0.2083&0.1543&3.7564&1\\
Staphylococcus epidermidis ATCC 12228&0.2277&0.2036&0.1399&3.7613&1\\
Staphylococcus haemolyticus JCSC1435&0.2304&0.2043&0.1526&3.7619&1\\
Streptococcus agalactiae 2603V/R&0.1690&0.1794&0.1200&3.8372&1\\
Streptococcus agalactiae NEM316&0.1679&0.1790&0.1209&3.8371&1\\
Streptococcus mutans UA159&0.1577&0.1783&0.1240&3.8468&1\\
Streptococcus pneumoniae R6&0.0952&0.1529&0.1210&3.9152&1\\
Streptococcus pneumoniae TIGR4&0.0957&0.1525&0.1209&3.9168&1\\
Streptococcus pyogenes M1 GAS&0.1227&0.1619&0.1137&3.8900&1\\
Streptococcus pyogenes MGAS10394&0.1167&0.1596&0.1101&3.8974&1\\
Streptococcus pyogenes MGAS315&0.1189&0.1636&0.1108&3.8929&1\\
Streptococcus pyogenes MGAS5005&0.1215&0.1612&0.1115&3.8929&1\\
Streptococcus pyogenes MGAS8232&0.1194&0.1608&0.1114&3.8932&1\\
Streptococcus pyogenes SSI-1&0.1189&0.1597&0.1111&3.8932&1\\
Streptococcus thermophilus CNRZ1066&0.1210&0.1710&0.1325&3.8908&1\\
Streptococcus thermophilus LMG 18311&0.1235&0.1737&0.1339&3.8881&1\\
Sulfolobus tokodaii str. 7&0.1932&0.1639&0.1253&3.7954&1\\
Thermoplasma acidophilum DSM 1728&0.0920&0.1668&0.2228&3.9315&2\\
Thermoplasma volcanium GSS1&0.0692&0.1345&0.1247&3.9379&2\\
Treponema polllidum str.Nichols&0.0548&0.0894&0.1095&4.0205&2\\
Ureaplasma parvun serovar 3 str. ATCC 700970&0.4111&0.2316&0.1950&3.5023&1\\
\hline
\end{longtable}
Thus, each genome is mapped into three-dimensional space determined by the indices (\ref{eq:1}~--
\ref{eq:4}). The Table provides also the fourth dimension, that is the absolute entropy of a codon
distribution. Further (see Section~\ref{classif}), we shall not take this dimension into consideration, since
it deteriorates the pattern observed in three-dimensional case.

Meanwhile, the data on absolute entropy calculation of the codon distribution for various bacterial genomes
are rather interesting. Keeping in mind, that maximal value of the entropy is equal to $S_{\max} = \ln 64 =
4.1589\ldots$, one sees that absolute entropy values observed over the set of genomes varies rather
significantly. {\sl Treponema polllidum str.Nichols} exhibits the maximal absolute entropy value equal to
$4.0205$, and {\sl Mycoplasma mycoides subsp. mycoides SC} has the minimal level of absolute entropy (equal
to $3.4356$).

\subsection{Classification}\label{classif}
\begin{figure}
\includegraphics[width=16cm]{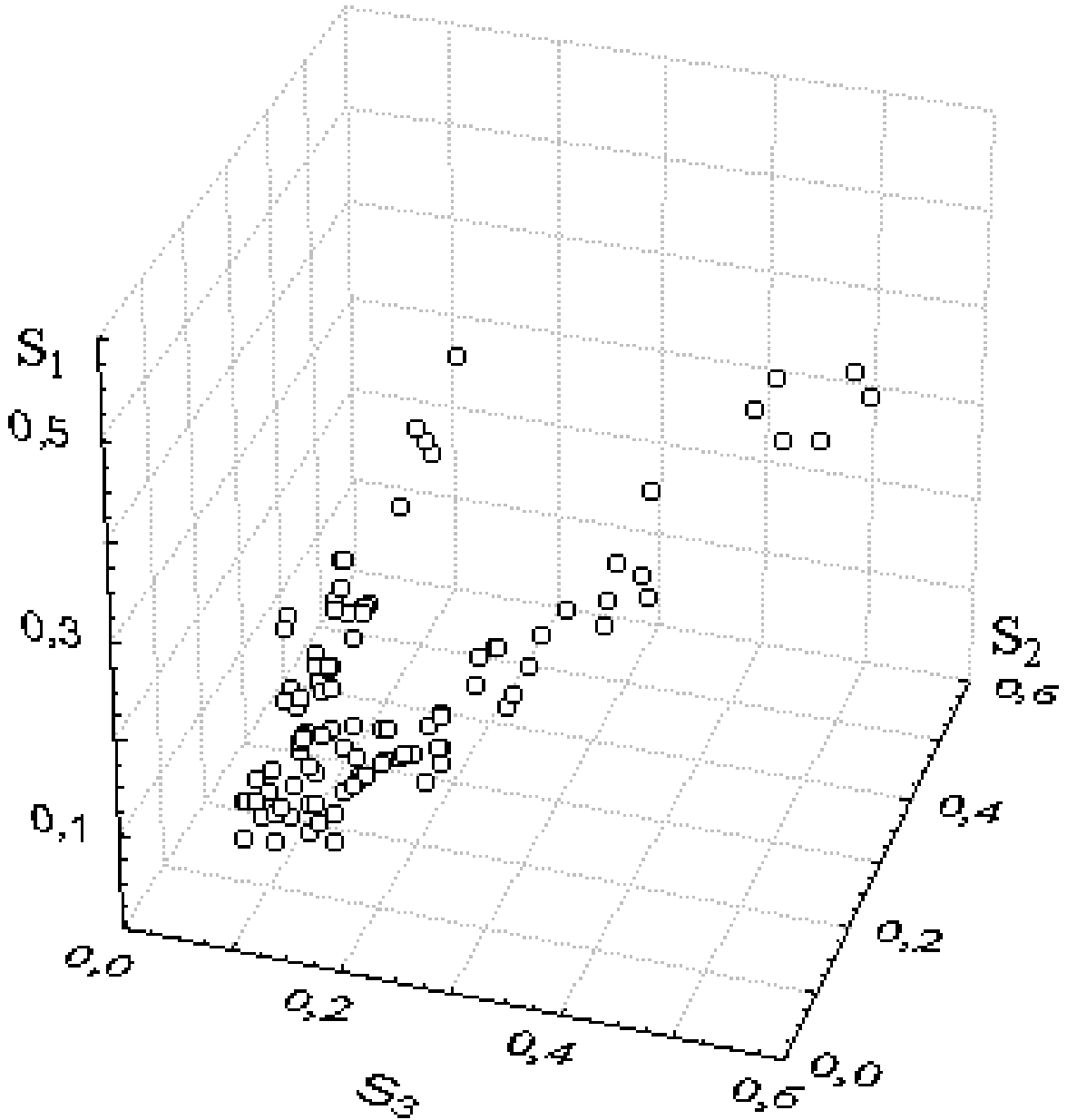}
\caption{\label{F1} The distribution of genomes in the space determined by the indices (\ref{eq:1}~--
\ref{eq:4}). $\mathsf{S}_1$~is $I$~based index, $\mathsf{S}_2$~is $S^{\ast}$~based index, and
$\mathsf{S}_3$~is $T$~based index of codon usage bias.}
\end{figure}

Consider a dispersion of the genomes at the space defined by the indices (\ref{eq:1}~-- \ref{eq:4}). The
scattering is shown in Figure~\ref{F1}. The dispersion pattern shown in this figure is two-horned; thus,
two-class pattern of the dispersion is hypothesized. Moreover, the genomes in the three-dimensional space
determined by the indices (\ref{eq:1}~-- \ref{eq:4}) occupy a nearly plane subspace. Obviously, the
dispersion of the genomes in the space is supposed to consists of two classes.

Whether the proximity of genomes observed at the space defined by three indices (\ref{eq:1}~-- \ref{eq:4})
meets a proximity in other sense, is the key question of our investigation. Taxonomy is the most natural idea
of proximity, for genomes. Thus, the question arises, whether the genomes closely located at the space
indices (\ref{eq:1}~-- \ref{eq:4}), belong the same or closely related taxons? To answer this question, we
developed an unsupervised classification of the genomes, in three-dimensional space determined by the indices
(\ref{eq:1}~-- \ref{eq:4}).

To develop such classification, one must split the genomes on $K$ classes, randomly. Then, for each class the
center is determined; that latter is the arithmetic mean of each coordinate corresponding to the specific
index. Then each genome (i.e., each point at the three-dimensional space) is checked for a proximity to each
$K$ classes. If a genome is closer to other class, than originally was attributed, then it must be
transferred to this class. As soon, as all the genomes are redistributed among the classes, the centers must
be recalculated, and all the genomes are checked again, for the proximity to their class; a redistribution
takes place, where necessary. This procedure runs till no one genome changes its class attribution. Then, the
discernibility of classes must be verified. There are various discernibility conditions (see, e.g.,
\citep{n5}).

Here we executed a simplified version of the unsupervised classification. First, we did not checked the class
discernibility; next, a center of a class differs from a regular one. A straight line at the space determined
by the indices (\ref{eq:1}~-- \ref{eq:4}) is supposed to be a center of a class, rather than a point in it.
So, the classification was developed with respect to these two issues. The Table~\ref{T1} also shows the
class attribution, for each genome (see the last column indicated as $C$).

\section{Discussion}\label{diskus}
Clear, concise and comprehensive investigation of the peculiarities of codon bias distribution may reveal
valuable and new knowledge towards the relation between the function (in general sense) and the structure of
nucleotide sequences. Indeed, here we studied the relation between the taxonomy of a genome bearer, and the
structure of that former. A structure may be defined in many ways, and here we explore the idea of ensemble
of (considerably short) fragments of a sequence. In particular, the structure here is understood in terms of
frequency dictionary (see Section~\ref{intro}; see also \citep{n1,n2,n3,n4} for details).

Figure~\ref{F1} shows the dispersion of genomes in three-dimensional space determined by the indices
(\ref{eq:1}~-- \ref{eq:4}). The projection shown in this Figure yields the most suitable view of the pattern;
a comprehensive study of the distribution pattern seen in various projections shows that it is located in a
plane (or close to a plane). Thus, the three indices (\ref{eq:1}~-- \ref{eq:4}) are not independent.

Next, the dispersion of the genomes in the indices (\ref{eq:1}~-- \ref{eq:4}) space is likely to hypothesize
the two-class distribution of the entities. Indeed, the unsupervised classification developed for the set of
genomes gets it. First of all, the genomes of the same genus belong the same class, as a rule. Some rare
exclusion of this rule result from a specific location of the entities within the ``bullet'' shown in
Figure~\ref{F1}.

A measure of codon usage bias is matter of study of many researchers (see, e.g., \citep{e3,e4,e5,e6,e8}).
There have been explored numerous approaches for the bias index implementation. Basically, such indices are
based either on the statistical or probabilistic features of codon frequency distribution \citep{1,2,e3},
others are based on the entropy calculation of the distribution \citep{3,x} or similar indices based on the
issues of multidimensional data analysis and visualization techniques \citep{e5,e5-1}. An implementation of
an index (of a set of indices) affects strongly the sense and meaning of the observed data; here the question
arises towards the similarity of the observations obtained through various indices implementation, and the
discretion of the fine peculiarities standing behind those indices.

Entropy seems to be the most universal and sustainable characteristics of a frequency distribution of any
nature \citep{bolz,obhod}. Thus, the entropy based approach to a study of codon usage bias seems to be the
most powerful. In particular, this approach was used by \cite{6}, where the entropy of the codon frequency
distribution has been calculated, for various genomes, and various fragments of genome. The data presented at
this paper manifest a significant correspondence to those shown above; here we take an advantage of the
general approach provided by \cite{6} through the calculation of more specific index, that is a mutual
entropy.

An implementation of an index (or indices) of codon usage bias is of a merit not itself, but when it brings a
new comprehension of biological issues standing behind. Some biological mechanisms affecting the codon usage
bias are rather well known \citep{e8,e4,2,e9,4,5}. The rate of translation processes are the key issue here.
Quantitatively, the codon usage bias manifests a significant correlation to $\mathsf{C}+\mathsf{G}$ content
of a genetic entity. Obviously, the $\mathsf{C}+\mathsf{G}$ content seems to be an important factor (see,
e.\,g. \citep{e5,e5-1}); some intriguing observation towards the correspondence between
$\mathsf{C}+\mathsf{G}$ content and the taxonomy of bacteria is considered in \citep{mist}.

Probably, the distribution of genomes as shown in Figure~\ref{F1} could result from $\mathsf{C}+\mathsf{G}$
content; yet, one may not exclude some other mechanisms and biological issues determining it. An exact and
reliable consideration of the relation between structure (that is the codon usage bias indices), and the
function encoded in a sequence is still obturated with the widest variety of the functions observed in
different sites of a sequence. Thus, a comprehensive study of such relation strongly require the
clarification and identification of the function to be considered as an entity. Moreover, one should provide
some additional efforts to prove an absence of interference between two (or more) functions encoded by the
sites.

A relation between the structure (that is the codon usage bias) and taxonomy seems to be less deteriorated
with a variety of features to be considered. Previously, a significant dependence between the triplet
composition of 16S\,RNA of bacteria and their taxonomy has been reported \citep{g1,g2}. We have pursued
similar approach here. We studied the correlation between the class determined by the proximity at the space
defined by the codon usage bias indices (\ref{eq:1}~-- \ref{eq:4}), and the taxonomy of bacterial genomes.

The data shown in Table~\ref{T1} reveal a significant correlation of class attribution to the taxonomy of
bacterial genomes. First of all, the correlation is the highest one for species and/or strain levels. Some
exclusion observed for {\sl Bacillus} genus may result from a modification of the unsupervised classification
implementation; on the other hand, the entities of that genus are spaced at the head of the bullet (see
Figure~\ref{F1}). A distribution of genomes over two classes looks rather complicated and quite irregular.
This fact may follow from a general situation with higher taxons disposition of bacteria.

Nevertheless, the introduced indices of codon usage bias provide a researcher with new tool for knowledge
retrieval concerning the relation between structure and function, and structure and taxonomy of the bearers
of genetic entities.

\section*{Acknowledgements} We are thankful to Professor Alexander Gorban from Liechester University for encouraging discussions of this work.

\end{document}